\documentclass[
twocolumn,
superscriptaddress,
 amsmath,amssymb,
 aps,
]{revtex4-1}
\usepackage{graphicx}
\usepackage{dcolumn}
\usepackage{bm}
\usepackage[USenglish]{babel}
\usepackage{xcolor}
\definecolor{prlblue}{rgb}{0.176, 0.152, 0.57}
\usepackage[colorlinks=true, linkcolor=black, citecolor=prlblue, filecolor=black, urlcolor=prlblue]{hyperref}
\usepackage{cleveref}
\begin{document}
\title{Controlled density-downramp injection in a beam-driven plasma wakefield accelerator}
\author{A.~Knetsch}
\email{alexander.knetsch@desy.de}
\affiliation{Deutsches Elektronen-Synchrotron DESY, Hamburg, Germany}
\author{B.~Sheeran}
\affiliation{Deutsches Elektronen-Synchrotron DESY, Hamburg, Germany}
\affiliation{University of Hamburg, Hamburg, Germany}
\author{L.~Boulton}
\affiliation{Deutsches Elektronen-Synchrotron DESY, Hamburg, Germany}
\affiliation{SUPA, Department of Physics, University of Strathclyde, Glasgow, UK}
\affiliation{The Cockcroft Institute, Daresbury, UK}
\author{P.~Niknejadi}
\author{K.~P\~oder}
\author{L.~Schaper}
\author{M.~Zeng}
\affiliation{Deutsches Elektronen-Synchrotron DESY, Hamburg, Germany}
\author{S.~Bohlen}
\affiliation{Deutsches Elektronen-Synchrotron DESY, Hamburg, Germany}
\affiliation{University of Hamburg, Hamburg, Germany}
\author{G.~Boyle}
\author{T.~Br\"ummer}
\affiliation{Deutsches Elektronen-Synchrotron DESY, Hamburg, Germany}
\author{J.~Chappell}
\affiliation{University College London, London, UK}
\author{R.~D'Arcy}
\affiliation{Deutsches Elektronen-Synchrotron DESY, Hamburg, Germany}
\author{S.~Diederichs}
\affiliation{Deutsches Elektronen-Synchrotron DESY, Hamburg, Germany}
\affiliation{University of Hamburg, Hamburg, Germany}
\author{B.~Foster}
\affiliation{John Adams Institute for Accelerator Science at University of Oxford, Oxford, UK}
\author{M.~J.~Garland}
\affiliation{Deutsches Elektronen-Synchrotron DESY, Hamburg, Germany}
\author{P.~Gonzalez~Caminal}
\affiliation{Deutsches Elektronen-Synchrotron DESY, Hamburg, Germany}
\affiliation{University of Hamburg, Hamburg, Germany}
\author{B.~Hidding}
\affiliation{SUPA, Department of Physics, University of Strathclyde, Glasgow, UK}
\affiliation{The Cockcroft Institute, Daresbury, UK}
\author{V.~Libov}
\author{C.~A.~Lindstr\o m}
\author{A.~Martinez~de~la~Ossa}
\affiliation{Deutsches Elektronen-Synchrotron DESY, Hamburg, Germany}
\author{M.~Meisel}
\affiliation{Deutsches Elektronen-Synchrotron DESY, Hamburg, Germany}
\affiliation{University of Hamburg, Hamburg, Germany}
\author{T.~Parikh}
\affiliation{Deutsches Elektronen-Synchrotron DESY, Hamburg, Germany}
\author{B.~Schmidt}
\affiliation{Deutsches Elektronen-Synchrotron DESY, Hamburg, Germany}
\author{S.~Schr\"oder}
\author{G.~Tauscher}
\affiliation{Deutsches Elektronen-Synchrotron DESY, Hamburg, Germany}
\affiliation{University of Hamburg, Hamburg, Germany}
\author{S.~Wesch}
\affiliation{Deutsches Elektronen-Synchrotron DESY, Hamburg, Germany}
\author{P.~Winkler}
\affiliation{Deutsches Elektronen-Synchrotron DESY, Hamburg, Germany}
\affiliation{University of Hamburg, Hamburg, Germany}
\author{J.~Wood}
\author{J.~Osterhoff}
\affiliation{Deutsches Elektronen-Synchrotron DESY, Hamburg, Germany}
\date{\today}

\begin{abstract}
This paper describes the utilization of beam-driven plasma wakefield acceleration to implement a high-quality plasma cathode via density-downramp injection in a short injector stage at the \textsc{FLASHForward} facility at DESY. Electron beams with charge of up to $105\,\mathrm{pC}$ and energy spread of a few percent were accelerated by a tunable effective accelerating field of up to $2.7\,\mathrm{GV/m}$. The plasma cathode was operated drift-free with very high injection efficiency. Sources of jitter, the emittance and divergence of the resulting beam were investigated and modelled, as were strategies for performance improvements that would further increase the wide-ranging applications for a plasma cathode with the demonstrated operational stability.
\end{abstract}
\maketitle

\section{introduction}
Plasma wakes excited by charged particle beams~\cite{rosenzweig1988experimental,chen1985AccElecPlasmaPRK,J.B.Rosenzweig1991AccelerationFields} are capable of generating GV/m accelerating gradients~\cite{TajimaDawson1979,Ruth1985AAccelerator}. This is 2\textendash3 orders of magnitude larger than conventional radio-frequency-based technologies, thereby 
promising a drastic reduction in both the size and cost of particle-accelerator facilities. Such a decrease is of particular significance for future linear colliders~\cite{Cros2019TowardsCollaboration} and the provision of compact free-electron laser (FEL) photon sources.  
The research field of beam-driven plasma-wakefield accelerators (PWFA) is dynamic, with advances ranging from the energy doubling of a $42\,\mathrm{GeV}$ electron beam over less than a meter~\cite{Blumenfeld2007EnergyAccelerator}, to the controlled correction of a correlated energy spread within a plasma channel of tens of millimeters \cite{DArcy2019TunableDechirper,Shpakov2019LongitudinalWakefields,Wu2019PhaseReduction}.
In particular, there has been significant progress in energy transfer from the beam driving the wake (the drive beam) to the trailing beam experiencing the accelerating field (the witness beam)~\cite{Litos2014High-efficiencyAccelerator,Loisch2018ObservationAcceleration,Corde2015Multi-gigaelectronvoltWakefield}. 
Witness beams can be injected into the plasma wakefield either from an external source or internally by the trapping of ambient plasma electrons. Research on external injection methods has predominantly concentrated on maximizing energy transfer from the drive to the witness beam while preserving other beam parameters, such as emittance and energy spread~\cite{Libov2018FLASHForwardBooster,Joshi2018PlasmaII}. Internal injection methods with extremely high electric-field gradients (GV/m) have the potential to generate beams with exceptionally high quality. As opposed to injecting a beam into an accelerating structure, in these methods, a relativistic electron beam with much smaller phase space is formed directly inside the accelerating structure. The predicted witness beams have normalized transverse emittance values much smaller than their drive beams and charges in the range of tens to hundreds of picocoulombs with femtosecond bunch durations~\cite{Hidding_PRL_2012,LiTransColliding2013,MartinezDeLaOssa2013High-qualityInjection}. Internal injection therefore offers the opportunity to 
generate a new class of beams with significantly enhanced brightness in comparison to conventional accelerator sources -- an extremely desirable feature for future photon sources and applications in high-energy physics.
While a variety of complementary internal injection methods have been shown to work in principle~\cite{Navid_distributed_PRL,Deng2019,Oz2007Ionization-inducedWakes}, precise control over the injection process, and consequently the injected witness-beam parameters, has so far remained relatively unexplored. 
This paper describes experiments carried out at \textsc{FLASHForward}~\cite{FlashForwardFacility2016,DArcy2019FLASHForward:Applications} \textendash~a dedicated plasma-wakefield beamline adjacent to the FLASH facility~\cite{Ackermann2007OperationWindow}, which operates with kiloampere-level beam currents and FEL-grade stability and beam quality~\cite{Schreiber2015TheFLASH}. It reports on the experimental demonstration of stable and controlled internal injection in PWFA utilizing a laser induced density downramp ~\cite{BulanovPRE98WaveBreak,suk2001_PRL_DensityTransition,Wittig2015OpticalAccelerators,Wittig2016ElectronSpikes} and demonstrates for the first time a plasma cathode that can be reliably operated. The injected bunch properties as a function of laser alignment and laser energy are also explored. 

\begin{figure*}[ht]
    \includegraphics[width=1\textwidth]{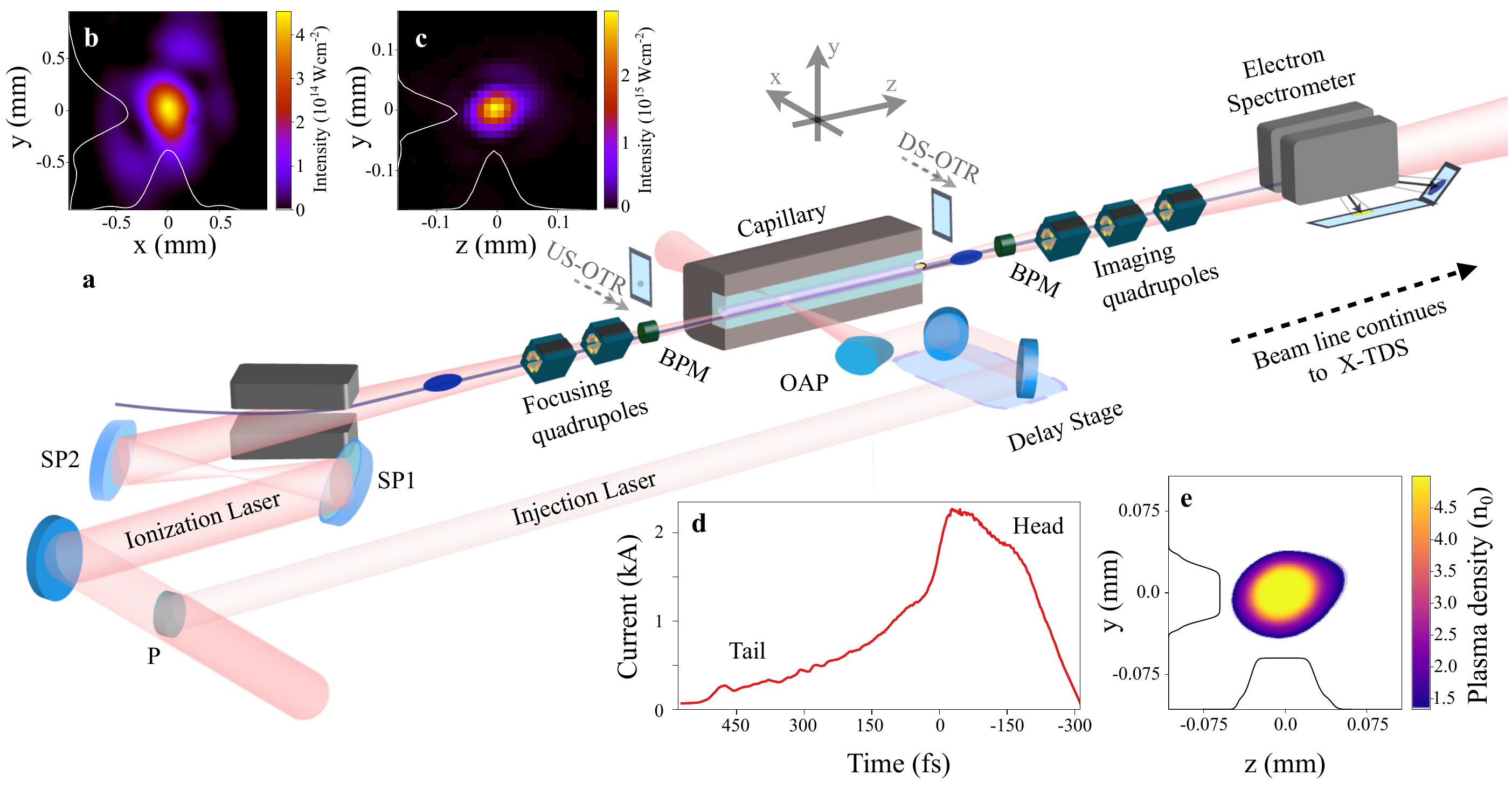}
    \caption{Schematic layout depicting the key elements of the experimental setup at \textsc{FLASHForward}. (a) Laser and electron beamline layout: P is the pick-off mirror for the transverse injection laser beam, SP1 and SP2 are spherical mirrors focusing the longitudinal ionization laser and OAP is the off-axis parabola focusing the injection laser. The e-beam position monitors (BPM) are shown; US-OTR and DS-OTR represent the upstream and downstream optical-transition-radiation screens, respectively. These can be moved into the beam path as alignment diagnostics. Foci for the (b) ionization and (c) injection lasers on the US-OTR are shown. A side view of the plasma density calculated in the injection region from (c) is displayed in (d) (normalized to the background plasma density $n_e$). Black lines in (d) and white lines in (b,c) are axial projections. (e) is the measured current profile of the drive beam.}
    \label{fig:Setup}
\end{figure*}
\section{Density Downramp injection}
Generation of high-quality electron beams in a plasma accelerator via internal injection requires the trapping of electrons in the accelerating phase of the plasma wake. These electron populations can originate from within the wake through ionization injection~\cite{Hidding_PRL_2012,MartinezDeLaOssa2013High-qualityInjection,Navid_distributed_PRL,Hidding2014UltrahighSources,Wan2016CollidingAccelerator} or from plasma electrons forming the boundary of the wake's ion cavity~\cite{BulanovPRE98WaveBreak,Thompson2004PlasmaSource}. Trapping will occur when electrons that enter the accelerating phase of the plasma wake reach or exceed the phase velocity of the wake itself ($v_\mathrm{e} \ge v_\phi$).
Therefore, electrons need either to enter the wakefield at sufficiently high velocity or the potential of the plasma wake needs to be deep enough to accelerate these electrons to $v_\phi$.
The velocity of the plasma electrons depends on the strength of the generated wakefield which in turn is related to the drive beam and plasma densities. In a plasma of constant density, $n_e$, the phase velocity is equal to the velocity of the relativistic drive beam $v_d$, i.e., close to that of light, $c$. Such a stringent condition readily precludes trapping of electrons from the plasma background unless the phase velocity can be locally reduced using a plasma-density gradient~\cite{Katsouleas1986PhysicalAccelerator,suk2001_PRL_DensityTransition,england2002PREDensityTransition}. 
A  plasma density that decreases along the direction of propagation of the driver leads to a longitudinal expansion of the plasma wake with a concurrently reduced phase velocity of
\begin{equation}
    v_\phi=v_d\left(\frac{1}{2n_e}\frac{\partial n_e(z)}{\partial z}\xi+1\right)^{-1},
    \label{eq:PhaseVelocity}
\end{equation}
where $\xi=z-ct$ is the longitudinal coordinate in the co-moving frame, $z$ is the longitudinal coordinate in the laboratory frame and $t$ is time.
As illustrated by Eq.~\ref{eq:PhaseVelocity}, the phase velocity on a density downramp decreases towards the back of the plasma wake, allowing fast plasma electrons to rephase into the accelerating region, where they eventually become trapped and can form a witness beam.
If trapping is avoided except on the density downramp, the injection process can be controlled and density-downramp injection characterised in detail.

Numerical studies with 3D particle-in-cell (PIC) codes predict that density-downramp injection with a sufficiently sharp gradient produces electron beams with transverse normalized emittances below $1\,\mathrm{\mu m}$~\cite{grebenyuk2014DownRamp,MartinezDeLaOssa2017,Xu2017HighRegime,Zhang2019EffectAccelerator}. Such sharp density downramps can be achieved either hydrodynamically, as commonly used in laser-driven plasma wakefield acceleration~\cite{gonsalves2011NatureTunableDensity,Faure2010InjectionChannel,schmid2010densityPRSTAB2010}, or by ionization of distinct gas species with two perpendicularly focused laser arms~\cite{Wittig2015OpticalAccelerators,Wittig2016ElectronSpikes}; the latter is the approach pursued in this work.
\section{Experimental setup}
\label{sec:Setup}
The experimental setup at \textsc{FLASHForward} is depicted in Fig.~\ref{fig:Setup}. Electron beams with $\sim 800\,\mathrm{pC}$ of charge reaching a mean energy of up to $1.25\,\mathrm{GeV}$ in the FLASH linear accelerator were used to drive a wake inside a capillary plasma source. The plasma  was created by ionization inside the capillary using two laser pulses \textendash~one focused along the electron-beam axis, the other transverse to it. Combined laser ionization thus formed a sharp density spike surrounded by a plateau region. Spatial alignment and synchronization were achieved with screens sensitive to optical transition radiation (OTR) and by measuring the plasma response due to beam-induced heating.
\subsection{The plasma source}
Constant-gas-flow capillaries have demonstrated excellent performance in plasma accelerators~\cite{gonsalves2011NatureTunableDensity,Gonsalves2019PetawattWaveguide,Butler2002GuidingWaveguide}, allowing for tailored, short injector stages.
For this experiment, a constant-flow capillary with a total length of $50\,\mathrm{mm}$ at a diameter of $1.5,\mathrm{mm}$  with two gas inlets was designed. The capillary incorporates an additional access port with~$300\,\mathrm{\mu m}$ diameter, used to couple in the transverse laser and located $20\,\mathrm{mm}$ downstream of the entrance, leaving a total maximum acceleration length of $30\,\mathrm{mm}$ for the injected witness beam. 
Gas-flow simulations performed with COMSOL Multiphysics~\cite{Dickinson2014COMSOLMini-review} show that the transverse port reduces the plateau gas density by less than~$2\,\%$ and therefore has a negligible effect on the gas-density profile.
Either pure argon, pure helium, or an arbitrary ratio of the two gases could be prepared in a mixing volume outside the central vacuum chamber before filling the capillary. In the experiments described in this work, the partial-pressure ratio between helium and argon was set to 2:1. Turbomolecular pumps were connected to the chamber to protect the beamline vacuum, reducing the ambient gas pressure by 3 to 4 orders of magnitude with respect to the pressure inside the capillary. A differential pumping system further reduced the pressure by 3 orders of magnitude both upstream and downstream of the central vacuum chamber. This design ensured that the quality of neither the drive nor witness beam was degraded by interfaces between vacuum chambers and scattering on ambient gas particles. 
\subsection{Laser ionization}

The gas mixture was ionized by two independently focused arms of a $25\,\mathrm{TW}$ Ti:sapphire laser system synchronized to the FLASH electron beam at the 100-femtosecond level~\cite{schulz2015femtosecondFLASH}. 
A $0.5\,\mathrm{'}$ pick-off mirror inserted into the laser path reflected the central part of the $34\,\mathrm{mm}$-diameter laser beam into the injection laser beamline. As depicted in Fig.~\ref{fig:Setup}(a), the remaining part of the laser continued to propagate along the ionization laser arm. 
A FWHM pulse duration of $(40\pm3)\,\mathrm{fs}$ was measured in the injection laser beamline using optimized spectral properties for a short pulse at the injection laser focus. Focusing of the ionization laser was achieved by two spherical mirrors with a resulting effective focal length of 18\,m. Such a large effective focal length produced a focal spot with a FWHM in the $x,y$-direction of $(333 \times 461)\,\mu\mathrm{m}$ (Fig.~\ref{fig:Setup}(b)).
As a result, the Rayleigh range of the spot approached the meter scale with an intensity profile capable of ionizing a plasma column sufficiently long and wide enough to span the entire length of the capillary whilst fully containing the plasma wake.
An $f/51$ off-axis parabola focused the injection laser perpendicular to the electron beam axis through a fused silica window. 
To minimize non-linear dispersion that can deteriorate the quality of the focal spot or pulse length, the window thickness was chosen to be  $1.5\,\mathrm{mm}$ only. 
Since the gas profile at the transverse port was assumed to be flat, the resulting plasma-density shape from ionization was defined by the gas-mixture ratio and the injection-laser focus with a FWHM spot size of $(57\times 48)\,\mu\mathrm{m}$ as shown in Fig.~\ref{fig:Setup}(c).
Since the intensity of the injection laser was set to be significantly higher than that of the ionization laser, it could ionize more strongly bound electrons from argon and helium via tunneling ionization. The resulting localized plasma-density spike, calculated from the focus measurement with the ADK model~\cite{ADK_original}, is shown in Fig.~\ref{fig:Setup}(d). 
The intensities of the ionization and injection lasers were set such that the former ionized argon up to its second level whilst the latter was capable of additionally ionizing the third ionization level of argon and the first of helium (see Table \ref{tab:IonizationEnergies}). With such distinct gas ionization, it was possible to control the shape of the plasma, which initially followed the intensity distribution of the laser foci and was subsequently expected to expand hydrodynamically on the nanosecond timescale~\cite{Durfee1993LightPulses,Shalloo2018HydrodynamicChannels}.
\begin{table}[ht]
    \begin{tabular}{c|c|c}
        Ion. level   & Ion. energy (eV) & I(W$_\mathrm{ADK}$=1 fs$^{-1}$) (W/cm$^2$)  \\
        \hline
        Ar-I   & 15.76 & $4.6\times10^{14}$\\
        Ar-II & 27.63 & $1.1\times10^{15}$ \\
        Ar-III & 40.74 & $2.2\times10^{15}$ \\
        He-I & 24.59 & $2.2\times10^{15}$ \\
        He-II  & 54.42 & $1.2\times10^{16}$ \\
        \hline
    \end{tabular}
    \caption{Relevant ionization energies~\cite{NIST_Ionization} and the corresponding laser intensities for a tunnel ionization rate~\cite{ADK_original} of 1~fs$^{-1}$.}
    \label{tab:IonizationEnergies}
\end{table}
\subsection{Alignment and synchronisation}
The ionization laser, injection laser, electron beam and capillary were required to be accurately spatially aligned. Furthermore, the laser timing was required to be set such that ionization occured shortly before the arrival of the electron drive beam. In order to minimize moving parts in the FLASH vacuum, the capillary and the upstream and downstream OTRs were statically mounted on one base plate and controlled in position and angle by a hexapod inside a separate vacuum chamber via a mechanical feedthrough~\cite{lens.org/179-396-520-666-03X}.
Two aluminum-coated silica screens were used to align the system by imaging the   optical transition radiation from the passing electron beam and light from the two lasers onto CCD cameras. These two reference points, 
$1.8\,\mathrm{cm}$ upstream of the capillary entrance and $27.4\,\mathrm{cm}$  downstream of the capillary exit, allowed for complete spatial overlap over the length of the plasma stage between the plasma generated by the ionization laser and the axis of the electron beam. The injection laser was aligned to the $y$-position of the electron-beam axis. The capillary was then moved such that each laser propagated through the appropriate channel in the capillary.
\begin{figure}[t]
    \centering
    \includegraphics[width=\columnwidth]{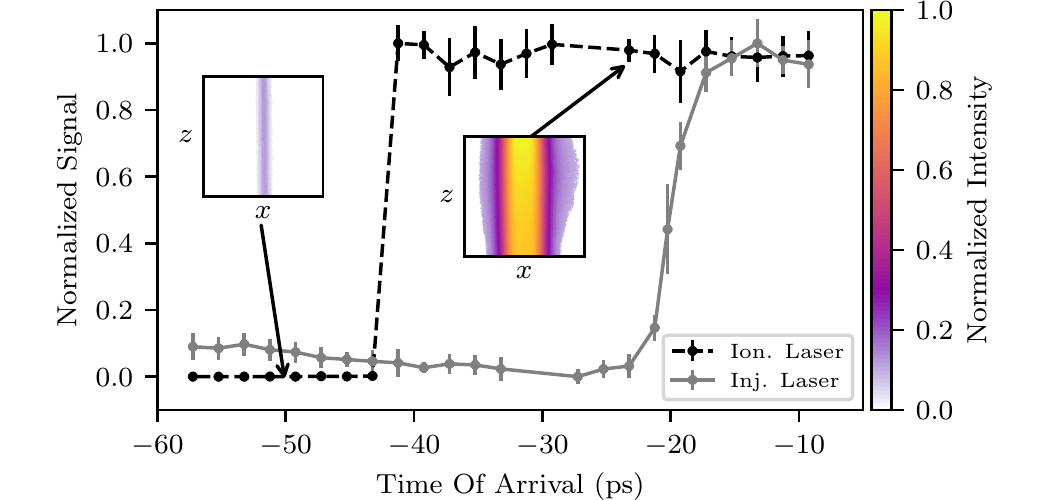}
    \caption{Normalized integrated plasma-afterglow signal as a function of the relative time of arrival (TOA) between the drive beam and the ionization (Ion.) and injection (Inj.) lasers; $t=0$ is defined as the TOA of the electron beam. Inset are examples of the signal enhancement due to electron-beam-induced heating of the plasma formed by the ionization laser.}
    \label{fig:plasmaglow_timing}
\end{figure}
Recent experimental results have shown that the recombination light from the plasma after the passage of an electron beam with currents of order kA can be used to synchronize lasers to electron beams to within a few femtoseconds~\cite{Scherkl2019Plasma-photonicBeams}.
This effect was used to synchronize the two laser arms to the drive beam. For this measurement, the plasma was ionized outside the capillary after filling the plasma chamber with argon gas to $0.05\,\mathrm{mbar}$. Figure \ref{fig:plasmaglow_timing} shows the time-of-arrival (TOA) dependence of plasma light collected by a CCD camera viewing from the top for each of the laser arms; examples of camera images are depicted as insets.
The visible transitions in Fig. 2 determine the time-of-arrival delay settings with respect to the electron beam of $42\,\mathrm{ps}$  for the ionization laser and $20\,\mathrm{ps}$ for the injection laser. Once the lasers were set up, the hexapod was positioned to maximise laser transmission through the capillary holes.

\subsection{The electron drive beam}
Downstream of the plasma, XQA quadrupoles~\cite{Okunev2016X-FELT/m} in combination with a spectrometer dipole and LANEX scintillating phosphor screens formed an imaging system capable of profiling electron beams from a few MeV to $2.4\,\mathrm{GeV}$. Alternatively, the electron beam could be transported to an X-band transverse deflecting structure~\cite{Marchetti2017X-BandProject,DArcy2018LongitudinalPolariX,DArcy2016AFlashforward} for precise longitudinal-phase-space characterisation. The current profile of the drive beam, calculated from such a measurement, is shown in Fig.~\ref{fig:Setup}(e). Additionally, the longitudinal phase space of the drive beam could be manipulated upstream of the plasma chamber with an energy collimator located in the dispersive section of the beamline~\cite{Schroder2020TunableFLASHForward}.
The charge and position of the drive beam were measured at various locations along the beamline by toroidal current transformers (toroids) in combination with stripline and cavity BPMs~\cite{Lipka2010DevelopmentXFEL}. 
The charge of the internally injected witness beam was determined in two independent ways: by subtracting the charges determined by a BPM directly downstream and a toroid upstream of the interaction point with a resolution of $1\,\mathrm{pC}$~\cite{Lipka2016FirstInjector}; and from a charge-calibrated phosphor screen in the electron spectrometer.
A drive beam with a charge of $(790\pm4)\,\mathrm{pC}$ was accelerated to $(1116\pm6)\,\mathrm{MeV}$ and compressed to a peak current of $2.1\,\mathrm{kA}$.
The beam, with a transverse root-mean-squared (rms) normalized emittance of $\epsilon_{x,y}=(14.0\times5.3)\,\mu\mathrm{m}$, was focused at the entrance of the capillary by four quadrupoles to a spot size of
$\sigma_{x,y}^\mathrm{d}=(25.5\pm1.6)\,\mathrm{\mu m}~\times~(17.1\pm0.8)\,\mathrm{\mu m}$, as measured by a transverse phase-space measurement using the combination of two BPMs and a scan of the spectrometer object plane~\cite{Lindstrm2020MatchingMonitors}. The plateau plasma density was determined to be $1.4 \substack{+0.4\\-0.3} \times 10^{16}~\mathrm{cm}^{-3}$ by comparing the measured and simulated energy losses of the drive beam, where the measured beam parameters, such as spot size, energy, current distribution and emittance were used as inputs to the simulation.
\section{Particle-In-Cell Simulations}
\label{sec:simulations}
The experiment was modeled with the 3D particle-in-cell (PIC) simulation code \textsc{OSIRIS}~\cite{Fonseca2002OSIRIS:Accelerators}.
The modeled plasma-density distribution with a plateau density of $1.4\times10^{16}\,\mathrm{cm}^{-3}$ was derived from gas-density profiles, simulated with COMSOL and tunneling ionization calculations~\cite{ADK_original}, based on the intensity distributions of the laser foci. Additionally, numerical studies with the code \textsc{FBPIC}~\cite{Lehe2016AAlgorithm} to model the pre-ionization showed that the longitudinal plasma-density profile beyond an acceleration length of 10 mm would develop a taper resulting from ionization defocussing. Such a taper can lead to a rephasing of the witness beam to lower accelerating fields. However, in this work, the discussion is limited to the injection process and the flat-top region; the longitudinal plasma-density profile that optimizes the acceleration to higher energies will be the subject of future work.

Simulations were conducted with a co-moving window of $(460\times400\times400)\,\mu\textrm{m}$ in $(z,x,y)$ with $(256\times256\times256)\,$cells. The drive-beam current profile, charge, and spot size were modeled based on the measurements described in Sec.~\ref{sec:Setup}. The number of  macro-particles per cell (PPC) for the drive beam was 8. While the PPC for the background plasma electrons was 1, the PPC was increased to 8 in a radius of $100\,\mathrm{\mu m}$ around the injection laser focus to resolve the physics of the injection process more accurately. 
The simulation case shown in this section represents the measurements shown in Figs.~\ref{fig:ONOFF}-\ref{fig:LaserEnergy_jitter} as well as the working point marked in Fig.~\ref{fig:OffsetScan}.
\begin{figure}[ht]
    \centering
    \includegraphics[width=\columnwidth]{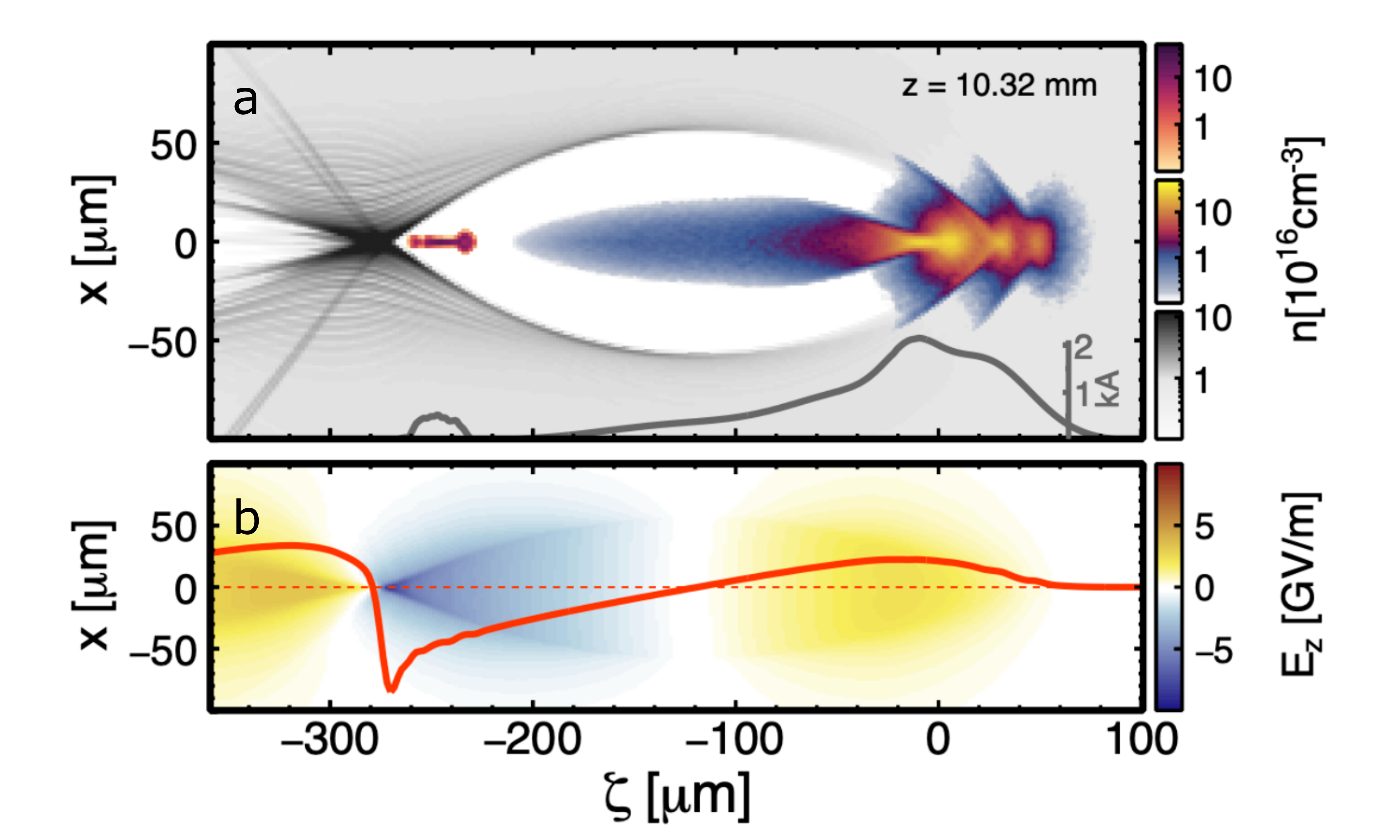}
    \caption{Simulation snapshot during the witness-beam acceleration:
    Cut through the plasma wake with electron densities (a) of witness beam (top right-hand colorbar), drive-beam (central colorbar), background plasma (bottom colorbar) and longitudinal current profile (solid gray line). (b) Longitudinal electric field as a central cut and an on-axis one-dimensional projection (solid red line).}
    \label{fig:Simulation_Snapshot}
\end{figure}

\begin{figure}[ht]
    \centering
    \includegraphics[width=\columnwidth]{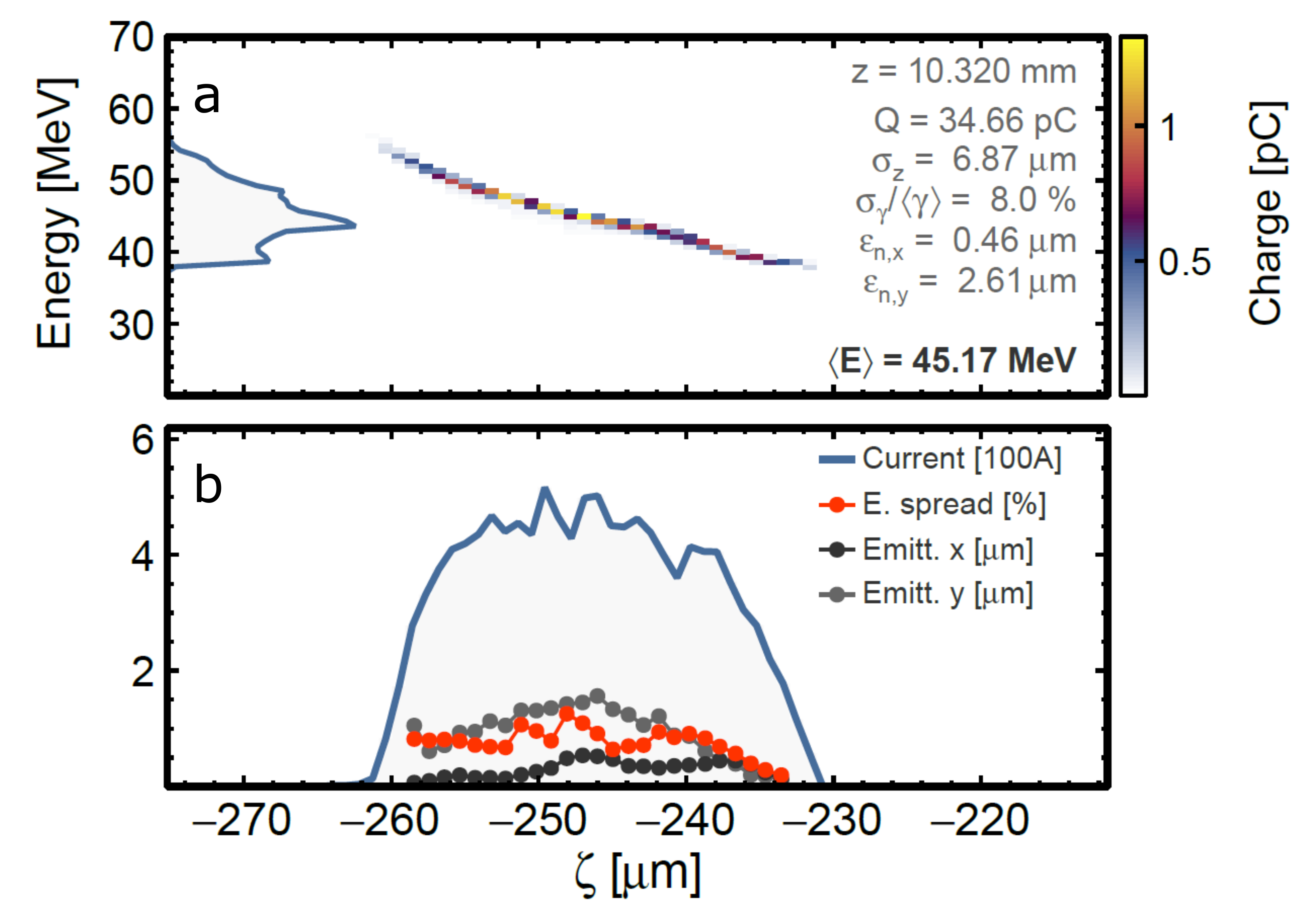}
    \caption{Longitudinal phase space of the simulated witness-beam parameter (a) and slice parameters (b) after $10.32\,\mathrm{mm}$ of acceleration.}
    \label{fig:Simulations_Witness}
\end{figure}

The slice properties and longitudinal phase space of the witness beam are shown in Fig.\,\ref{fig:Simulations_Witness}.
The simulated witness-beam charge is in excellent agreement with the measured values discussed in Sec.~\ref{sec:Results}. 

As described in Sec.~\ref{sec:Results}, the injection laser was positioned $(23.5\pm 4.5)\,\mathrm{\mu m}$ off axis with respect to the electron beam for the majority of the data taking.
The simulations show that such an offset leads to an increased emittance in the $y$ direction compared to sub-$\mu$m emittance in $x$. In future experiments, this asymmetry can be avoided with wider transverse foci, enabling sub-$\mu$m emittance beams in both planes, as predicted for similar parameters~\cite{MartinezDeLaOssa2017}.

Figure~\ref{fig:Simulations_Witness} shows that the simulated witness beam has a strongly correlated energy spread. While relatively large energy spread is a characteristic but undesirable property of electron beams from many plasma accelerators, a linearly correlated longitudinal phase space such as that shown in Fig.~\ref{fig:Simulations_Witness}(a)
can be corrected for in a downstream plasma dechirper stage~\cite{DArcy2019TunableDechirper,Shpakov2019LongitudinalWakefields,Wu2019PhaseReduction,FerranPousa2019CompactCompensation} or by overloading the plasma wake with additionally injected charge~\cite{ManahanDechirper2017}.
\section{Experimental Results}
\label{sec:Results}
\subsection{Laser-controlled injection}
\label{sec:Stable_Injection}
\begin{figure}[h]
    \centering
    \includegraphics[width=\columnwidth]{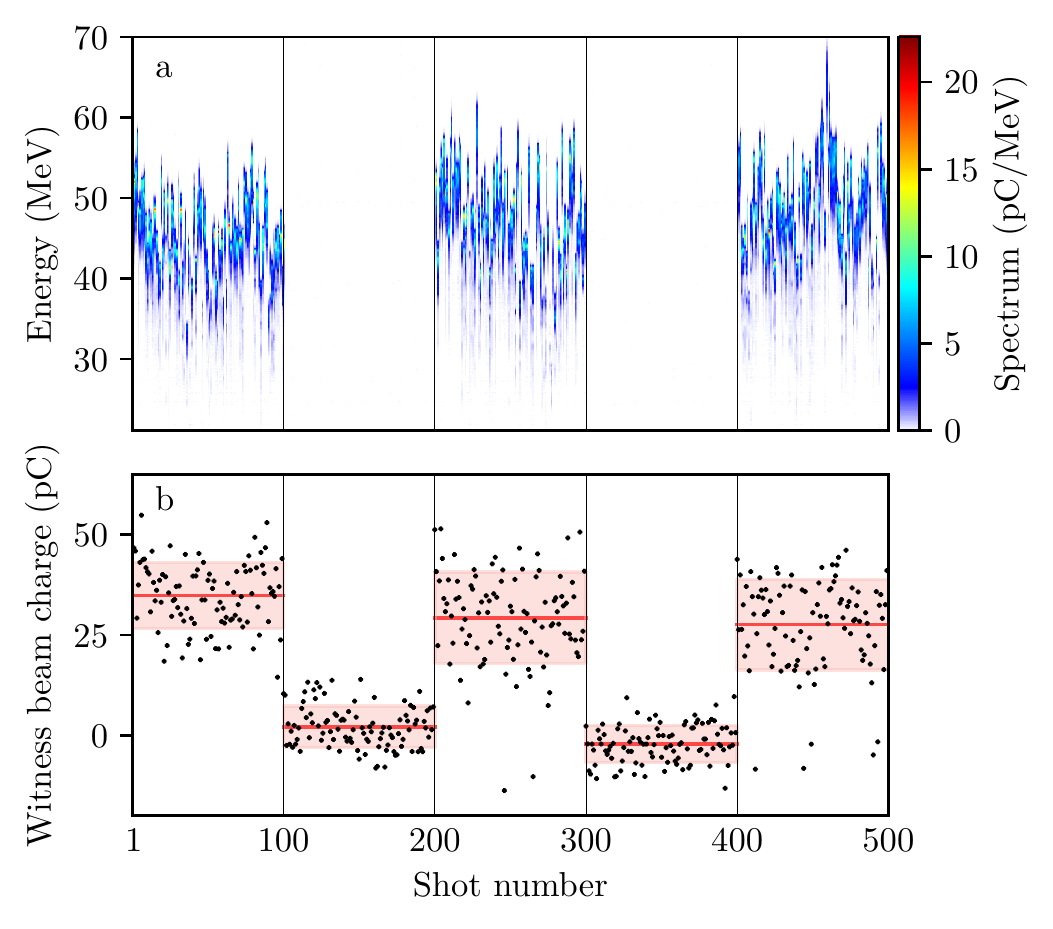}
    \caption{(a) Energy spectra and (b) charge of the injected witness beam. 
    The injection is switched off by inserting a beam block inside the injection-laser path for shot numbers 101--200 and 301--400. The mean value for each group of 100 shots is indicated by a solid line, while the rms variation of charge values is shown by a shaded area.}
    \label{fig:ONOFF}
\end{figure}

The first investigation explored how the witness beams depended on the presence of the injection laser. This is shown in Fig.~\ref{fig:ONOFF}.
Panel~(a) depicts the energy spectrum of the witness beam over 500 consecutive shots at a spectrometer imaging energy of ${45}\,\mathrm{MeV}$, while the corresponding excess-charge values are plotted in panel~(b). The injection laser was blocked and unblocked for 100 shots at a time. There is a 100\% correlation between the existence of a witness beam and the presence of the injection laser, with no entries being outside the range of the plot.
The excess charge was $Q_\mathrm{on}=(32.1\pm9.6)\,\mathrm{pC}$ when the injection laser was switched on and $Q_\mathrm{off}=(0\pm5)\,\mathrm{pC}$ otherwise. 
The injection process is therefore unambiguously triggered by the injection laser. Furthermore, the data confirm that there is no contribution from the ionization laser, in agreement with the PIC simulations (see Sec.\,\ref{sec:simulations}), so that downramp injection is established as the sole mechanism for witness-beam generation.
\subsection{Stability of witness-beam parameters}
\begin{figure}[h]
    \centering
    \includegraphics[width=\columnwidth]{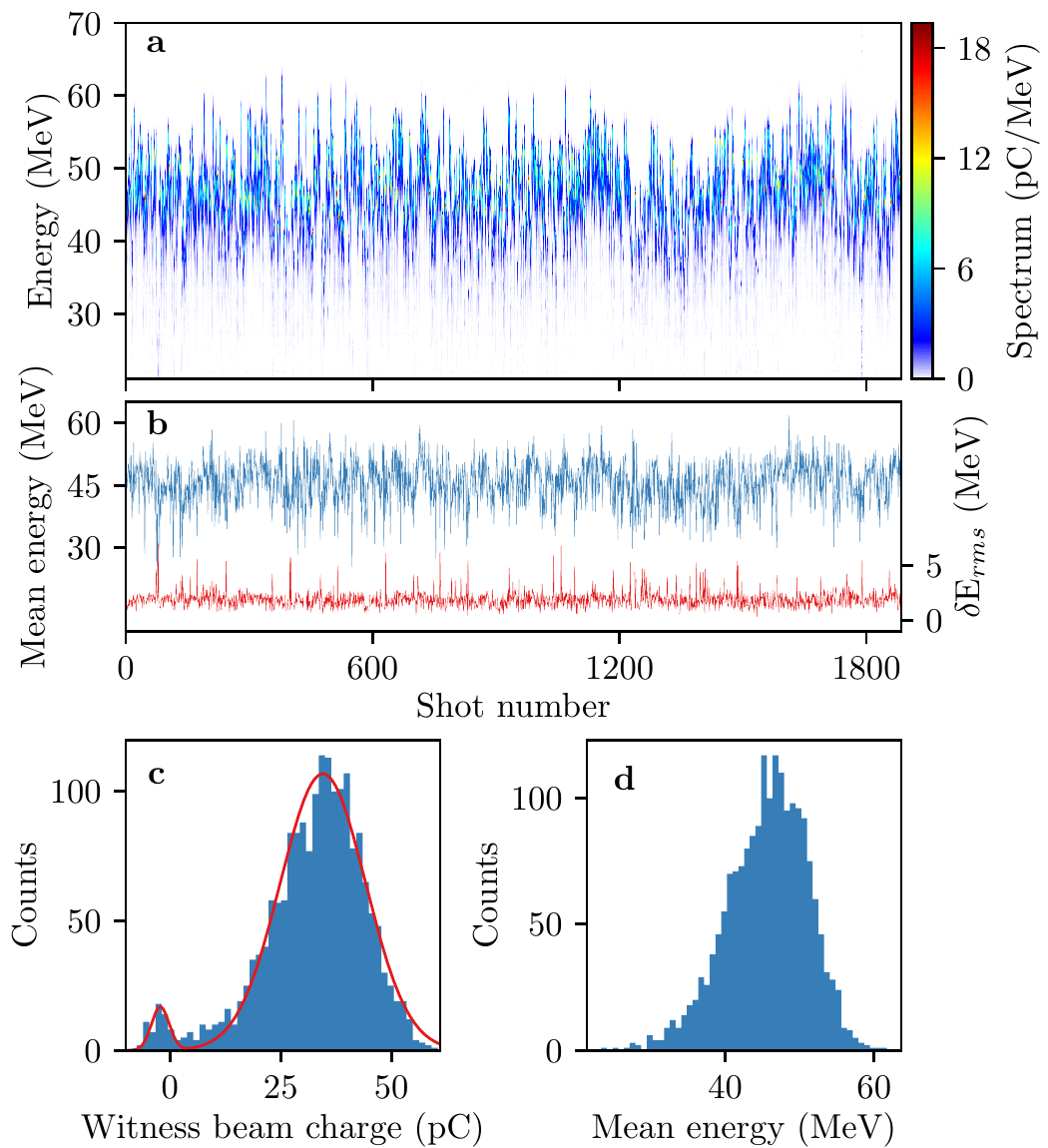}
    \caption{(a) Waterfall plot of the witness-beam spectrum for 1,885 consecutive shots. (b) The electron-beam mean energy (blue) and rms energy spread (red). (c) Histogram of the witness-beam charge, measured from the calibrated spectrometer screen with a double-Gaussian fit shown in red. (d) Histogram of the witness beam mean energy. Each histogram has 50 bins.}
    \label{fig:stability_run}
\end{figure}
A dataset of 1,885 consecutive shots taken at a repetition rate of 2\,Hz and recorded over a time span of approximately~15~minutes was used to explore the reliability of the plasma cathode.
A waterfall plot of the projected witness-beam energy spectra is shown in Fig.~\ref{fig:stability_run}(a). The spectra are remarkably stable. Fig.~\ref{fig:stability_run}(b) shows the evolution of the mean energy and  rms energy spread.
An injected witness beam was measured for $(95.4 \pm 2.5){\%}$ of the shots presented in Fig.~\ref{fig:stability_run}(a). 
The witness beam had a mean charge of $(33\pm10)\,\mathrm{pC}$ and mean energy of $(45\pm 5)\,\mathrm{MeV}$ with an rms relative energy spread of $ \langle \frac{\delta E}{E} \rangle =4.4\,\%$. 
\subsection{Witness-beam emittance and divergence}
\begin{figure}[h]
    \centering
    \includegraphics[width=\columnwidth]{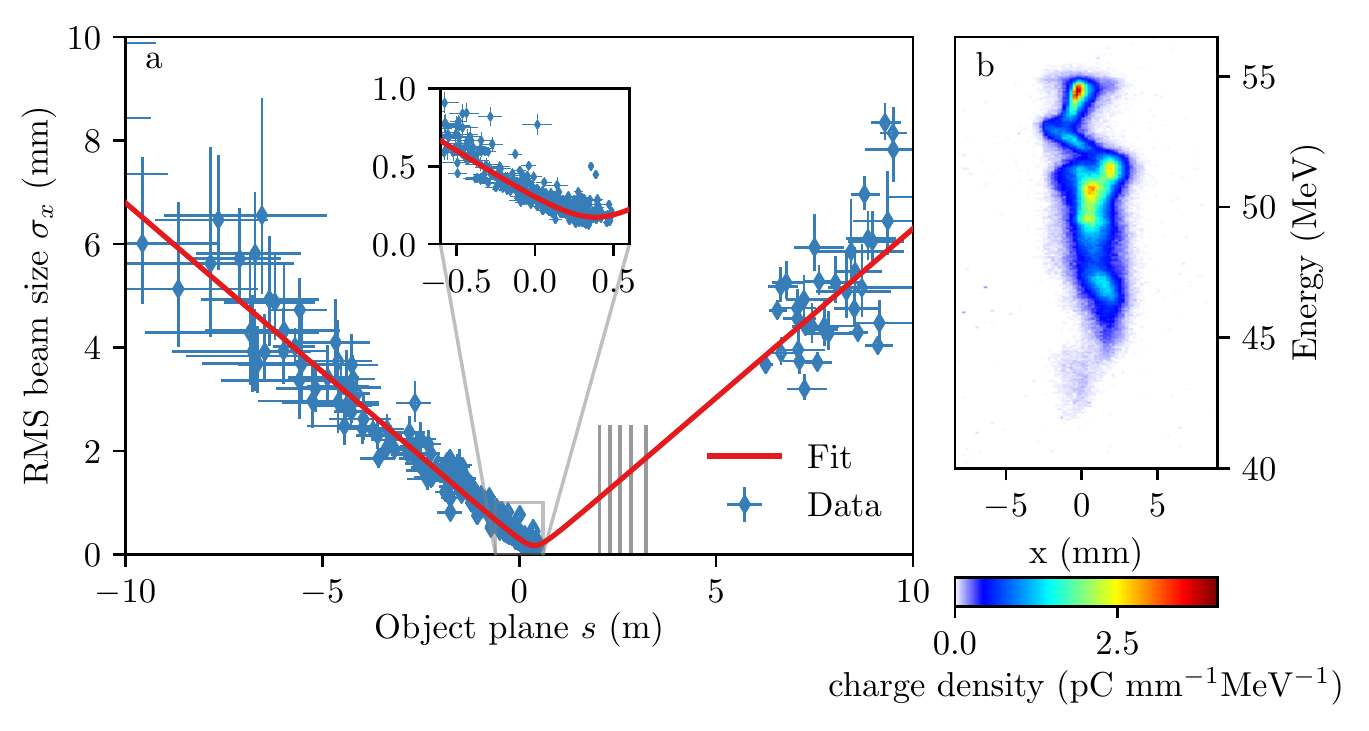}
    \caption{(a) Scan of the imaging quadrupole focusing strength.
  Each data point represents a single shot with an injected witness beam.
Transverse beam sizes on the spectrometer screen and mean Lorentz factor $\langle\gamma\rangle$ were calculated from the spectrometer signal.
From the quadrupole strength and $\langle\gamma\rangle$, the object plane $s$ and the magnification were calculated to extract $\sigma_x(s)$ at the corresponding object plane.
Error bars on $\sigma_x$ and $s$ were calculated analogously based on the rms-variation of the witness-beam spectral projection. The exit of the capillary is at $s=0$. Object planes in the range between $s= 0.2\,\mathrm{m}$~\textendash~$5.9\,\mathrm{m}$ lie too close to the quadrupoles (shown as gray lines) to be imaged.(b) An example of a witness-beam signal as measured on the spectrometer screen.}
    \label{fig:emittance}
\end{figure}
In the absence of focusing forces, a witness beam with a mean Lorentz factor $\langle\gamma\rangle$, transverse normalized emittance $\varepsilon_x^n$, and a beta function $\beta^*$ at its virtual source point $s_0$ will expand outside the plasma such that it has a transverse rms beam size of
\begin{equation}
    \sigma_x(s)=\sqrt{\frac{\varepsilon_x^n}{\langle\gamma\rangle} \left(\beta^* +\frac{(s-s_0)^2}{\beta^*}\right)}.
    \label{eq:Sigma_x}
\end{equation} at longitudinal position $s$.

Due to the stability of the injection process, this free-space propagation can be examined in a multi-shot measurement by imaging witness beams from different object-plane positions, $s$. 
Figure~\ref{fig:emittance}(a) shows the results of a scan of the quadrupole focusing strength of the imaging spectrometer, giving transverse beam size at the corresponding object plane, which gives information about the transverse expansion of the beam.
Fitting Eq.~\ref{eq:Sigma_x} to the data gives a divergence of $\theta=(0.65 \pm 0.01)\,\mathrm{mrad}$, $s_0=(381 \pm 6)\,\mathrm{mm}$, and a transverse rms normalized emittance projected over all shots of $\varepsilon_x^n=(9.3\pm 0.3)\,\mathrm{\mu m}$.
The fact that $s_0$ appears to be downstream of the capillary exit is probably an effect of the gas-density transition between the exit of the capillary and the vacuum of the plasma chamber. 
According to hydrodynamic simulations, gas-density ramps with atomic densities of order $10^{14}\,\text{cm}^{-3}$ can extend over tens of centimeters. This gas would be ionized by the ionization laser due to its long Rayleigh length. Plasma wakes driven in such a low-density plasma would continue to focus the witness beam for several centimeters beyond the capillary, possibly explaining the particularly low divergence. 

An example of the witness-beam signal measured on the spectrometer screen is shown in Fig.~\ref{fig:emittance}(b). Since the longitudinal phase space of a witness beam in a plasma accelerator is typically predominantly linear (see Sec.~\ref{sec:simulations}), the visible oscillatory structure can be interpreted as transverse centroid oscillations of longitudinal witness-beam slices. This is likely to contribute to the relatively high measured emittance compared to predicted values from numerical studies.
\subsection{Influence of laser energy on the plasma cathode}
\label{sec:LaserEnergy}
Understanding the influences on the witness-beam characteristics is of primary importance in controlling the injection process whilst optimising its stability, a prerequisite for a reliable electron-beam source.
As the plasma is laser ionized, variations in laser-pulse properties are important.
No dependence of the charge or energy of the witness beam on laser timing was detected up to at least $\approx 200\,\mathrm{ps}$ relative time-of-arrival between the drive-beam and the laser pulses. 
The ionization and the injection laser arms were derived from the same laser system, so the inherent laser-energy jitter may affect both the longitudinal and transverse laser ionization and thus the shape of both plasmas.
\begin{figure}[t]
    \centering
    \includegraphics[width=\columnwidth]{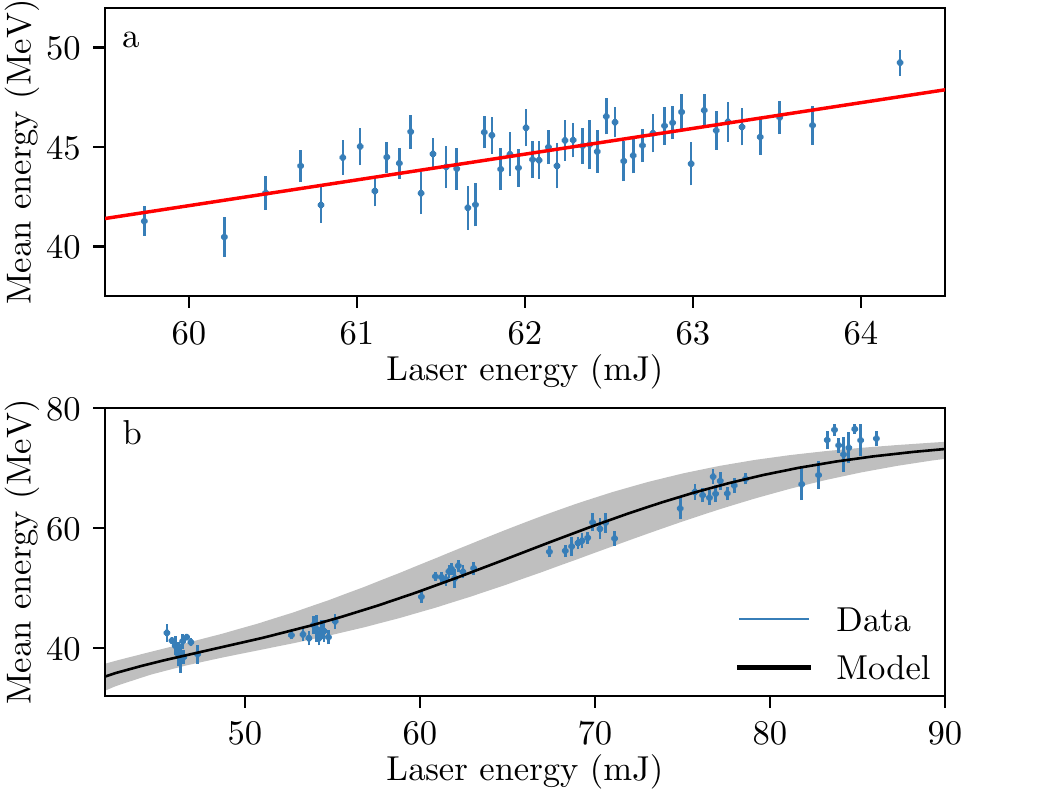}
    \caption{(a) Correlation between witness beam energy and total laser energy for the dataset shown in Fig.~\ref{fig:stability_run}(a). The line is the result of a straight-line fit to the data points. (b) Results from a laser energy scan, highlighting the variation of witness-beam energy with on-target laser energy. The solid line is a fit based on the model of Eq.~\ref{eq:model}; the gray area marks the uncertainty of the model caused by the pulse-length jitter.}
    \label{fig:LaserEnergy_jitter}
\end{figure}
The effect of the laser energy stability on the witness-beam energy was studied by examining the dataset of 1,885 consecutive shots on which Fig.~\ref{fig:stability_run} was based. The variation of mean witness-beam energy as a function of on-target laser energy is plotted in Fig.~\ref{fig:LaserEnergy_jitter}(a). The observed linear correlation shows that the witness-beam energy jitter will depend on the stability of the laser energy.
This relationship is illustrated more clearly by systematically varying the injection-laser energy over a much larger range, between $45\,\mathrm{mJ}$ and $86\,\mathrm{mJ}$. The data is shown in
Fig.~\ref{fig:LaserEnergy_jitter}(b). In the fit, a model is used which is based on the fact that higher plasma densities lead to higher accelerating fields and consequently to higher witness-beam energies. Since the plasma-density changes are relatively small in this energy range, the accelerating field  acting upon the witness beam $E_\mathrm{acc}$ was assumed to be proportional to the wave-breaking field $E_\mathrm{WB} = c   \sqrt{\frac{n_e m_e}{\epsilon_0}}$ \cite{Dawson1959NonlinearPlasma}.
The witness-beam energy $\mathcal{E}=e L_\mathrm{acc} E_\mathrm{acc}$ can be parametrized as
\begin{equation}
    \label{eq:model}
  \mathcal{E}= \eta e L_\mathrm{acc} E_\mathrm{WB}= \eta  e L_\mathrm{acc}c \sqrt{\frac{m_e }{\epsilon_0}[n_{e}(U_\mathrm{laser})+n_\mathrm{offset}]},
\end{equation}
where $U_\mathrm{laser}$ is the on-target laser energy, $\eta$ is a constant adjusting for deviations in effective acceleration length and deviations from acceleration at wave-breaking field-strength and $n_\mathrm{offset}$ describes an offset in effective average plasma density.
The values of $n_e(U_\mathrm{laser})$ were calculated numerically using the ADK model~\cite{ADK_original}, assuming that the laser pulse length and spot size were constant inside the plasma. 

The success of this model is illustrated in Fig.~\ref{fig:LaserEnergy_jitter}(b), showing that the change in the mean energy of the witness beam with the total laser energy can be understood as an increase in the average plasma density due to additional Ar-II ionization at higher laser energies.
The fit to the data results in 
$n_\mathrm{offset}= (-65.5 \pm 1.7)\times10^{14}\,\mathrm{cm}^{-3}$ and $\eta = 0.206 \pm 0.002$, indicating that the witness beam experiences a lower average plasma density than expected for a uniform longitudinal plasma profile and acceleration at a non-ideal phase or reduced acceleration length.
These findings can be explained, for example, with a tapered longitudinal profile caused by ionization defocussing as discussed in Section~\ref{sec:simulations}. 

The laser rms energy jitter was measured to be $1.5\,\%$. The slope extracted from the model described above translates this into an rms energy jitter of the witness beam of $3.9\,\%$.  
This is sufficiently small that  the accelerating field can be fine-tuned between ~1.3 and 2.7~GV/m by varying the laser energy. Control can be improved further by either designing  laser systems with lower energy jitter or by using different gas mixtures. For example, if the argon in the gas mixture were to be replaced with hydrogen, the ionization stemming from the ionization laser can saturate, reducing the sensitivity to the laser energy.
\subsection{Influence of the injection laser position on the plasma cathode}
\label{sec:OffsetScan}
\begin{figure}[ht]
    \centering
    \includegraphics[width=\columnwidth]{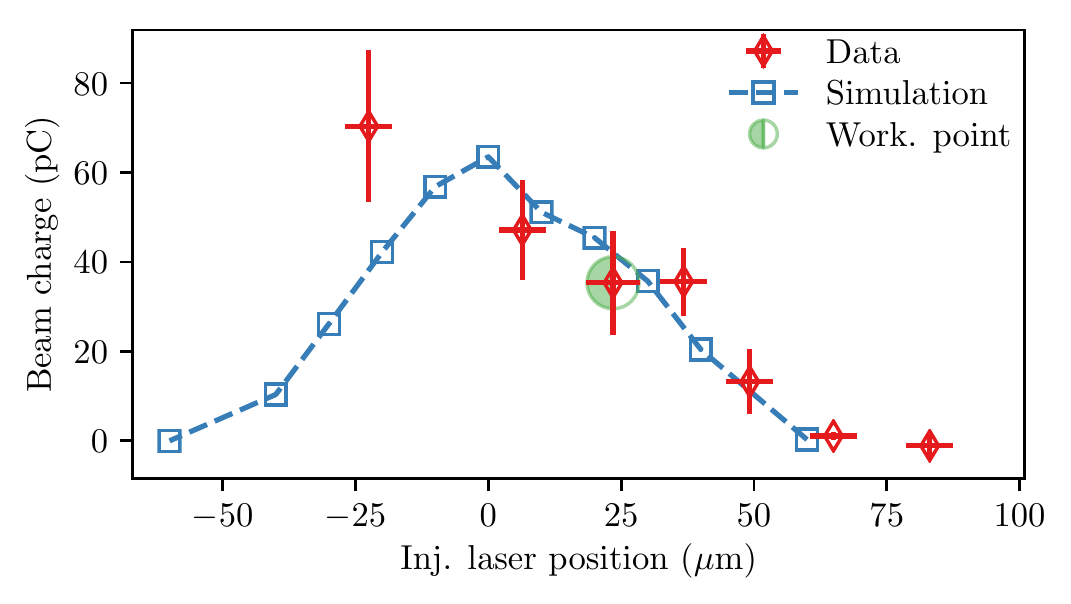}
    \caption{Effect of centroid laser-spot positioning with respect to the electron beam. The laser position was measured on the upstream OTR camera and the charge on the electron spectrometer screen (points with error bars).
    Results are compared to simulation results, shown as squares (for details see Sec.~\ref{sec:simulations}), with the plasma profile modeled as shown in Fig.~\ref{fig:Setup}(d). The working point for the data presented in this paper is indicated by the half-filled circle.}
    \label{fig:OffsetScan}
\end{figure}

That precise alignment between the two laser arms and the electron beam is crucial to stable injection can be seen in 
Fig.~\ref{fig:OffsetScan}, which shows the effect on the injected witness charge of changing the height of the injection laser with respect to the electron-beam axis. The height was changed by moving the optical assembly of the injection laser in the $y$-direction along the laser-beam path such that the alignment with respect to the off-axis parabola remained unchanged.
The relative calibration between focus position and motor position was carried out with the upstream OTR camera.

Every data point plotted in Fig.~\ref{fig:OffsetScan} represents the average of 20 events; the error on the measured charge is given by the rms variation of the witness-beam charge.
The positioning error bar, $\sigma_y^{e,\mathrm{inj}}=4.5\,\mathrm{\mu m}$, was constant throughout the dataset and was calculated from two main contributions: the rms $y$-position jitter of the electron beam at the injection position, $\sigma_y^e=1.4\,\mathrm{\mu m}$, measured by two cavity BPMs~\cite{Lipka2010DevelopmentXFEL} around the plasma chamber; and the rms variation of the laser centroid $y$ position, $\sigma_{y}^\mathrm{inj}=4.2\,\mathrm{\mu m}$, measured at the focus.
The influence of the jitter on the $z$ position of the injection laser, $\sigma_{z}^\mathrm{inj}=3.4\,\mathrm{\mu m}$ was neglected, as it led to a difference in longitudinal injection position that made a negligible contribution to the witness-beam energy jitter. Furthermore, the $x$-position jitter of the electron-beam waist, $\sigma_{x}^e=1.0\,\mathrm{\mu m}$ was also assumed to have a negligible effect, since it was much smaller than the Rayleigh length of the injection-laser focus. The scan was only performed on one side of the injection-laser position distribution.

A comparison between the experimental data and a simulated offset scan is plotted in Fig.~\ref{fig:OffsetScan}, based on the measured plasma profile (see Fig.~\ref{fig:Setup}(d)). In this Figure, zero is at the highest simulated injected charge. Since the relative offset between the electron beam and injection laser was not measured to sub-spot-size accuracy, it was determined by fitting  the simulation-based model to the data. The data presented in Figs.~\ref{fig:ONOFF} - \ref{fig:LaserEnergy_jitter} were taken at the point indicated in Fig.~\ref{fig:OffsetScan} as the ``work. point", which shows lower sensitivity to the relative alignment than the experimentally determined position of maximum charge at $y=(-22.5 \pm 4.5)\,\mathrm{\mu m}$.
The slope of the simulation-based model gives an estimate of the charge variation resulting from position jitter, $\delta Q_\mathrm{wit.}(\sigma_{y}^{e,\mathrm{inj}})= 6.9\,\mathrm{pC}$
The charge variation in the witness bunch can be explained predominantly by the position jitter between the electron beam and the injection laser, which is known to be dominated by the laser pointing jitter. This jitter would be eliminated by using an injection laser with a spot size that is  many times wider than the plasma wake. Such a shape can be produced with focusing optics that produce asymmetric foci, such as cylindrical lenses. 

\section{Summary}
A plasma cathode with stability at a level unprecedented in the field of beam-driven plasma injectors has been demonstrated, studied, and operated reliably.  Effective accelerating gradients of GV/m were demonstrated, tunable in the range of 1.3 -- 2.7~GV/m. The stability allowed a multi-shot measurement of the beam-divergence and emittance to be performed. Sources of charge and energy jitter were identified and mitigation strategies were proposed for future applications. Particle-in-cell simulations gave good agreement with the observed witness-beam parameters. These indicate that the requirements for future compact FELs of sub-micron emittances in both planes, peak currents of a few hundred amperes and narrow energy spread can be attained using the methods described in this paper. These results constitute a significant step towards stable, controllable plasma-based cathodes and brightness converter stages, which are of great interest for next-generation photon-science and particle-physics facilities.
\begin{acknowledgments}
The authors would like to thank M. Dinter, S. Karstensen, K. Ludwig, F. Marutzky, A. Rahali, V. Rybnikov, A. Schleiermacher and S. Thiele, as well as the FLASH accelerator team, DESY MVS and the DESY FH and M divisions
for their engineering and technical support.
We thank the OSIRIS consortium (IST/UCLA) for access to the OSIRIS code and acknowledge the use of the High-Performance Cluster (Maxwell) at DESY. We also gratefully acknowledge the Gauss Centre for Supercomputing e.V. (www.gauss-centre.eu) for funding this project by providing computing time through the John von Neumann Institute for Computing (NIC) on the GCS Supercomputer JUWELS at J{\"u}lich Supercomputing Centre (JSC).
B.H. and L.B. were supported by the European Research Council (ERC) under the European Union’s Horizon 2020 research and innovation programme (NeXource, ERC Grant agreement No. 865877).
We furthermore thank the developers of the pipeline-analysis software, Manuel Kirchen and S\"oren Jalas from the University of Hamburg.
\end{acknowledgments}

\end{document}